\documentclass[12pt]{article}

\usepackage{amsmath,amssymb,cite,bm}
\usepackage{graphicx}

\makeatletter
 
  \@addtoreset{equation}{section}
 \makeatother

\newcommand{\n}{\nonumber \\}
\newcommand{\Tr}{\mathrm{Tr}}

\newcommand{\diag}{\mathrm{diag}}

\newcommand{\cL}{\mathcal{L}}

\newcommand{\be}{\begin{equation}}
\newcommand{\ee}{\end{equation}}
\newcommand{\bea}{\begin{eqnarray}}
\newcommand{\eea}{\end{eqnarray}}
\newcommand{\beann}{\begin{eqnarray*}}
\newcommand{\eeann}{\end{eqnarray*}}

\newcommand{\ba}{\begin{array}}
\newcommand{\ea}{\end{array}}

\allowdisplaybreaks

\begin{document}

\setlength{\oddsidemargin}{0cm}
\setlength{\baselineskip}{7mm}

\begin{titlepage}
\renewcommand{\thefootnote}{\fnsymbol{footnote}}
\begin{normalsize}
\begin{flushright}
\begin{tabular}{l}
KUNS-2501 \\
YITP-14-46 \\
KEK-TH-1740
\end{tabular}
\end{flushright}
  \end{normalsize}

~~\\

\vspace*{0cm}
    \begin{Large}
       \begin{center}
         {Emergent bubbling geometries in \\ gauge theories with $SU(2|4)$ symmetry}
       \end{center}
    \end{Large}
\vspace{0.7cm}

\begin{center}
Yuhma A{\sc sano}$^{1)}$\footnote
            {
e-mail address : 
yuhma@gauge.scphys.kyoto-u.ac.jp }, 
Goro I{\sc shiki}$^{\phantom{(}2)}$\footnote
            {
e-mail address : 
ishiki@yukawa.kyoto-u.ac.jp}, 
    {\sc and}
Shinji S{\sc himasaki}$^{\phantom{(}1),\phantom{(}3)}$\footnote
           {
e-mail address : 
simasaki@post.kek.jp }\\
      
\vspace{0.7cm}
                    
     $^{1)}$ {\it Department of Physics, Kyoto University}\\
               {\it Kyoto, 606-8502, Japan}\\

     $^{2)}$ {\it Yukawa Institute for Theoretical Physics, Kyoto University}\\
               {\it Kyoto, 606-8502, Japan}\\

     $^{3)}$ {\it Theory Center, High Energy Accelerator Research Organization (KEK)}\\
               {\it Ibaraki 305-0801, Japan}\\

\end{center}

\vspace{0.7cm}

\begin{abstract}\noindent
We study the gauge/gravity duality between bubbling geometries in type IIA supergravity
and gauge theories with $SU(2|4)$ symmetry, 
which consist of $\mathcal{N}=4$ super Yang-Mills on $R\times S^3/Z_k$,
$\mathcal{N}=8$ super Yang-Mills on $R\times S^2$ and the plane wave matrix model.
We show that the geometries are realized as field configurations in the strong coupling region 
of the gauge theories.
On the gravity side, the bubbling geometries can be mapped to electrostatic systems with conducting disks.
We derive integral equations which determine the charge densities on the disks.
On the gauge theory side, we obtain a matrix integral by applying the localization 
to a 1/4-BPS sector of the gauge theories.
The eigenvalue densities of the matrix integral turn out to satisfy the same integral equations as 
the charge densities on the gravity side.
Thus we find that these two objects are equivalent.

\end{abstract}
\vfill

\end{titlepage}
\vfil\eject

\setcounter{footnote}{0}


\section{Introduction}
Emergent geometry is a key concept in formulating a quantum theory of gravity because 
in such a theory space(time) is believed to be not fundamental but emergent.
A concrete realization of this notion has been achieved by the gauge/string duality \cite{Maldacena,GKP,Witten},
which states an equivalence between gauge theories and string theories.
For instance, the duality between the $c=1$ matrix model and the two-dimensional string theory
 is known to be a prototype of the gauge/string duality 
(see for reviews \cite{Klebanov:1991qa,Ginsparg:1993is,Mukhi:2003sz}). 
In this duality, the matrix model can be mapped to a free fermion system.
The Fermi surface, which is the eigenvalue density of the $c=1$ matrix model, corresponds to 
one spatial direction of the dual two-dimensional geometry.
Another well-known example is 
the duality between 1/2-BPS operators in $\mathcal{N}=4$ super Yang-Mills theory (SYM) and the bubbling geometries 
in type IIB supergravity \cite{LLM,Yamaguchi:2006te}\footnote{
The duality for less supersymmetric operators such as 1/4- and 1/8-BPS were studied, for example, in \cite{Chen:2007du}.
}.
The chiral primary operators in $\mathcal{N}=4$ SYM can also be mapped to the states of free fermions 
\cite{Berenstein:2004kk,Takayama:2005yq} and
its phase space can be identified with the droplets in the gravity dual \cite{LLM}.
The same kind of correspondence is shown to exist for the 1/2-BPS Wilson loop operators in \cite{Yamaguchi:2006te}.
In these cases also, the Fermi surface corresponds to one spatial direction in the dual ten-dimensional geometry.
Other recent examples include the duality between Wilson surfaces in six-dimensional $\mathcal{N}=(2,0)$ theory 
and M-theory bubbling geometries \cite{Mori:2014tca}.

Recently, yet another example of the emergent geometry 
has been demonstrated by the present authors and T. Okada in \cite{Asano:2014vba}.
There, we studied the gauge/gravity duality between the plane wave matrix model (PWMM)
and the bubbling geometry in type IIA supergravity \cite{LLM,LM}, which is explained in detail below.

PWMM is a matrix quantum mechanics originally proposed as a formulation of M-theory  in the light-cone frame
on the pp-wave geometry \cite{BMN}. 
The theory is the mass deformation of the BFSS matrix theory \cite{BFSS} and
 has $SU(2|4)$ symmetry, which consists of $R\times SO(3)\times SO(6)$ bosonic symmetry\footnote{
Precisely speaking, this is the bosonic subgroup of the universal cover of $SU(2|4)$.} 
and 16 supersymmetries.
PWMM possesses many discrete vacua, called fuzzy sphere, which are labeled by  representations of $SU(2)$.
One can make the irreducible decomposition of the representations, 
so that the vacua are specified by irreducible representations and their multiplicities 
which appear in the decomposition.

For each vacuum of PWMM, the dual gravity solution was constructed by Lin and Maldacena in type IIA supergravity, 
which is known as the bubbling geometry \cite{LLM,LM}.
If $SU(2|4)$ symmetric ansatz is assumed, the equations of motion can be reduced to a simple differential equation 
with boundary conditions characterized by fermionic droplets
on a particular one-dimensional line in ten-dimensions.
Solving this equation is shown to be equivalent to solving for an electrostatic potential
of a three-dimensional axially symmetric electrostatic system with some conducting disks \cite{LM}.
In this duality, the dimensions of irreducible representations of a fuzzy sphere vacuum 
correspond to NS5-brane charges and their multiplicities correspond to D2-brane charges.
In the associated electrostatic system,
they correspond to the positions and the charges of the conducting disks.
See Fig.1 for an electrostatic system for a general vacuum of PWMM.

In \cite{Asano:2014vba}, the duality for PWMM was studied in the case of the vacuum characterized 
by the direct sum of copies of an irreducible representation.
The gravity dual of this theory was examined in \cite{Ling:2006up},
where the charge density on the conducting disk in the corresponding electrostatic system 
was shown to satisfy a Fredholm integral equation of the second kind.
In PWMM,  a 1/4-BPS sector was considered. This sector is made of a complex scalar
 that corresponds to two spatial directions in the dual geometry on which the electrostatic system is defined.
By applying the localization technique \cite{Pestun:2007rz}, 
this sector reduces to a matrix integral \cite{Asano:2012zt}.
By evaluating this matrix integral, it was found that
the eigenvalue density of the matrix integral obeys the same integral equation as the charge density 
in the electrostatic problem. 
This fact naturally leads us to identify the eigenvalue density with the charge density.
Since the charge density completely determines the gravity solution,
it was concluded that one can exactly reconstruct the gravity solution from PWMM.

Remarkably, the sector studied in \cite{Asano:2014vba} is interacting and so 
the coupling constant appears as a parameter in this duality.
This is in sharp contrast to the examples given in the first paragraph, 
where the relevant sectors  in the gauge theories are free and there is no coupling dependence.
The existence of such an extra parameter makes the duality more attractive. 
Actually, the gauge/gravity duality for PWMM admits two interesting limits 
that lead to the duality for other field theories \cite{Maldacena:2002rb,Ling:2006up}.
One is the D2-brane limit (commutative limit of fuzzy sphere), in which PWMM becomes SYM on $R\times S^2$, 
and the dual geometry asymptotically becomes D2-brane geometry.
The other one is the NS5-brane limit, in which 
PWMM is considered to become type IIA little string theory (LST) 
\cite{Berkooz:1997cq,Seiberg:1997zk,Aharony:1999ks,Kutasov:2001uf} 
on $R\times S^5$,
and the dual geometry asymptotically becomes NS5-brane geometry.
Thus, the study of the gauge/gravity duality for PWMM enables us to study that
for SYM on $R\times S^2$ and even that for LST in a comprehensive manner.

In this paper, we extend the results of \cite{Asano:2014vba} to the case of a general vacuum of PWMM.
We first consider the gravity side. 
We  solve the electrostatic problem for a general vacuum of PWMM
by extending the method developed in \cite{Ling:2006up}. 
We then show that the electrostatic potential of the problem
can be described in terms of the charge densities on each conducting disk
and the charge densities satisfy a system of integral equations.
Next, we study the gauge theory side.
We consider a general vacuum and analyze the 1/4-BPS sector as in the previous paper \cite{Asano:2014vba}.
In this case, the 1/4-BPS sector can be described by a multi-matrix integral \cite{Asano:2012zt}.
We show that the saddle point equations for the eigenvalue densities of the matrix integral 
are exactly the same as the system of integral equations for the charge densities 
in the corresponding electrostatic problem. 
Thus, as in \cite{Asano:2014vba}, 
we can identify the eigenvalue densities with the charge densities, which determine the gravity solution.
This result shows that PWMM around a general vacuum can contain its gravity dual geometry 
as saddle point configurations of eigenvalues.

Moreover, we investigate the gauge/gravity duality for other gauge theories with $SU(2|4)$ symmetry,
$\mathcal{N}=8$ SYM on $R\times S^2$ and $\mathcal{N}=4$ SYM on $R\times S^3/Z_k$.
It was shown in \cite{LM,Ishiki:2006yr} that the gauge/gravity duality for SYM on $R\times S^2$
can be obtained from PWMM by taking the D2-brane limit 
and that for SYM on $R\times S^3/Z_k$ by taking the T-duality as well as the D2-brane limit.
Using these relations, we prove the correspondence between eigenvalue densities in these theories
and charge densities in the electrostatic systems associated with their gravity duals.

The organization of this paper is as follows.
In section 2, we review the dual gravity solutions for gauge theories with $SU(2|4)$ symmetry 
and solve general electrostatic problems associated to these solutions.
In section 3, after reviewing PWMM and the matrix integral, 
we derive the saddle point equations for the eigenvalue densities.
Then, we show that the charge densities can be identified with the eigenvalue densities.
Finally, we discuss the cases for SYM on $R\times S^2$ and SYM on $R\times S^3/Z_k$.

\section{Gravity dual of gauge theories with $SU(2|4)$ symmetry}

In this section, we elaborate the Lin-Maldacena solution for a general vacuum of the gauge theories 
with $SU(2|4)$ symmetry by analyzing the corresponding electrostatic system.
We will show that the charge densities on each disk, which determine the gravity solution, 
satisfy a system of integral equations.

\subsection{Lin-Maldacena solution}
First, we review the Lin-Maldacena solution \cite{LM}, 
which is the solution with $SU(2|4)$ symmetry in type IIA supergravity. 
The solution is given by
\begin{align}
ds_{10}^2 &= 
\left( \frac{\ddot{V}-2\dot{V}}{-V''} \right)^{1/2}
\left\{-4 \frac{\ddot{V}}{\ddot{V}-2\dot{V}}dt^2
-2 \frac{V''}{\dot{V}}(dr^2 +dz^2)
+4 d\Omega_{5}^2 +2 \frac{V'' \dot{V}}{\Delta} d\Omega_2^2
\right\}, \nonumber\\
C_1 &= - \frac{(\dot{V}^2)'}{\ddot{V}-2\dot{V}} dt, \;\;\;
C_3 = -4 \frac{\dot{V}^2 V''}{\Delta} dt \wedge d\Omega_2 , 
\nonumber\\
B_2 &= \left( 
\frac{(\dot{V}^2)'}{\Delta}+2z \right) d\Omega_2, \;\;\;
e^{4\Phi} = \frac{4(\ddot{V}-2\dot{V})^3}{-V'' \dot{V}^2 \Delta^2},
\label{LM solution}
\end{align}
where $\Delta = (\ddot{V}-2\dot{V})V''-(\dot{V}')^2$ and 
the dots and primes denote $\frac{\partial}{\partial \log r} $
and $\frac{\partial }{\partial z}$, respectively.
A remarkable feature of this solution is that 
it is written in terms of a single function $V(r,z)$.
The Killing spinor equation in the supergravity imposes a condition 
that $V(r,z)$ satisfies the Laplace equation in a three-dimensional
axially symmetric electrostatic system, where $r$ and $z$ represent 
coordinates for the transverse and the axial directions, respectively.
The regularity of the metric requires that 
the electrostatic system must consist of some conducting disks with radii tuned such that
the charge densities vanish at the edges.
In addition, from the positivity of the metric, there must be a certain background potential.
So, the potential $V(r,z)$ consists of these two contributions:
\begin{align}
V(r,z)=V_{\rm b.g.}(r,z)+ \tilde{V}(r,z).
\label{Vbg+tildeV}
\end{align}

\begin{figure}[tbp]
\begin{center}
\includegraphics[height=7cm, keepaspectratio, clip]{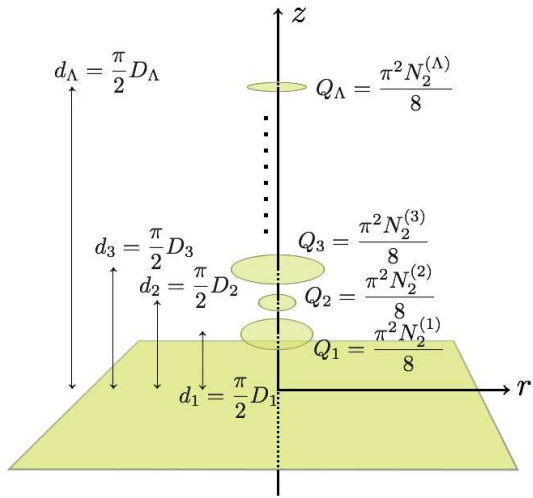}
\end{center}
\caption{Electrostatic system corresponding to a general vacuum of PWMM.}
\end{figure}

The electrostatic system is determined once a theory and its vacuum are specified.
The electrostatic system relevant to PWMM consists of 
an infinite conducting plate at $z=0$, some finite conducting disks in the region of $z \geq 0$ (Fig.1)
and the background potential of the form
\begin{align}
V_{\rm b.g.}(r,z)= V_0 \left(r^2z -\frac{2}{3}z^3\right),
\label{bg pot}
\end{align}
where $V_0$ is a constant.
The electrostatic system relevant to $\mathcal{N}=8$ SYM on $R\times S^2$  consists of 
some finite conducting disks in the region $-\infty\leq z\leq \infty$ (Fig.2) 
and the background potential of the form
\begin{align}
V_{\rm b.g.}(r,z)=W_0(r^2-2z^2),
\label{bg pot S2}
\end{align}
where $W_0$ is a constant.
The electrostatic system relevant to $\mathcal{N}=4$ SYM on $R\times S^3/Z_k$ consists of
an infinite number of finite conducting disks arranged periodically along the $z$-axis (Fig.3)
and the background potential \eqref{bg pot S2}\footnote{
Note that \eqref{bg pot S2} is periodic up to terms of linear and constant in $z$, 
which do not contribute to the gravity solution \eqref{LM solution}.}.

\begin{figure}[tbp]
\begin{center}
\includegraphics[height=7cm, keepaspectratio, clip]{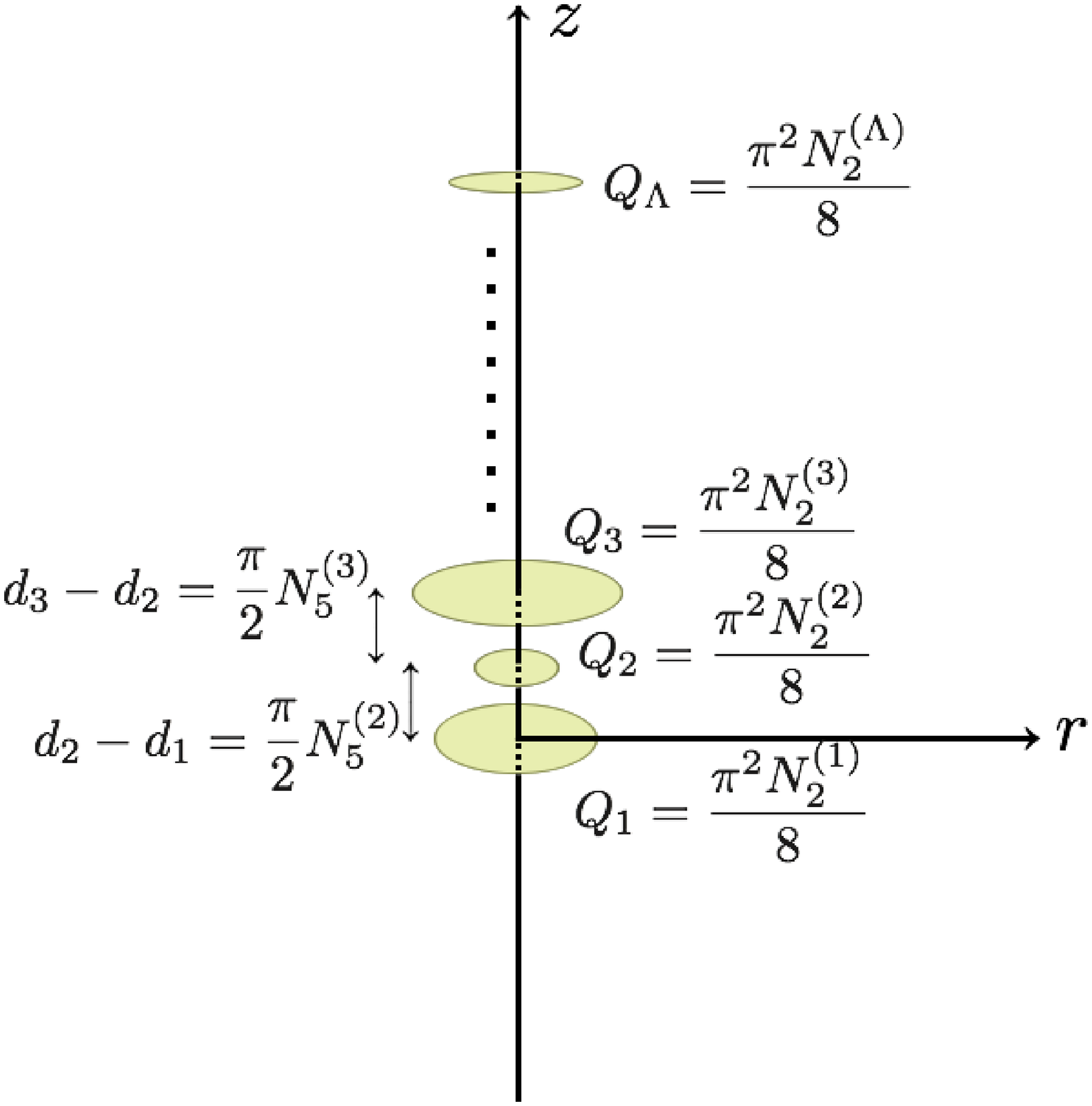}
\includegraphics[height=7cm, keepaspectratio, clip]{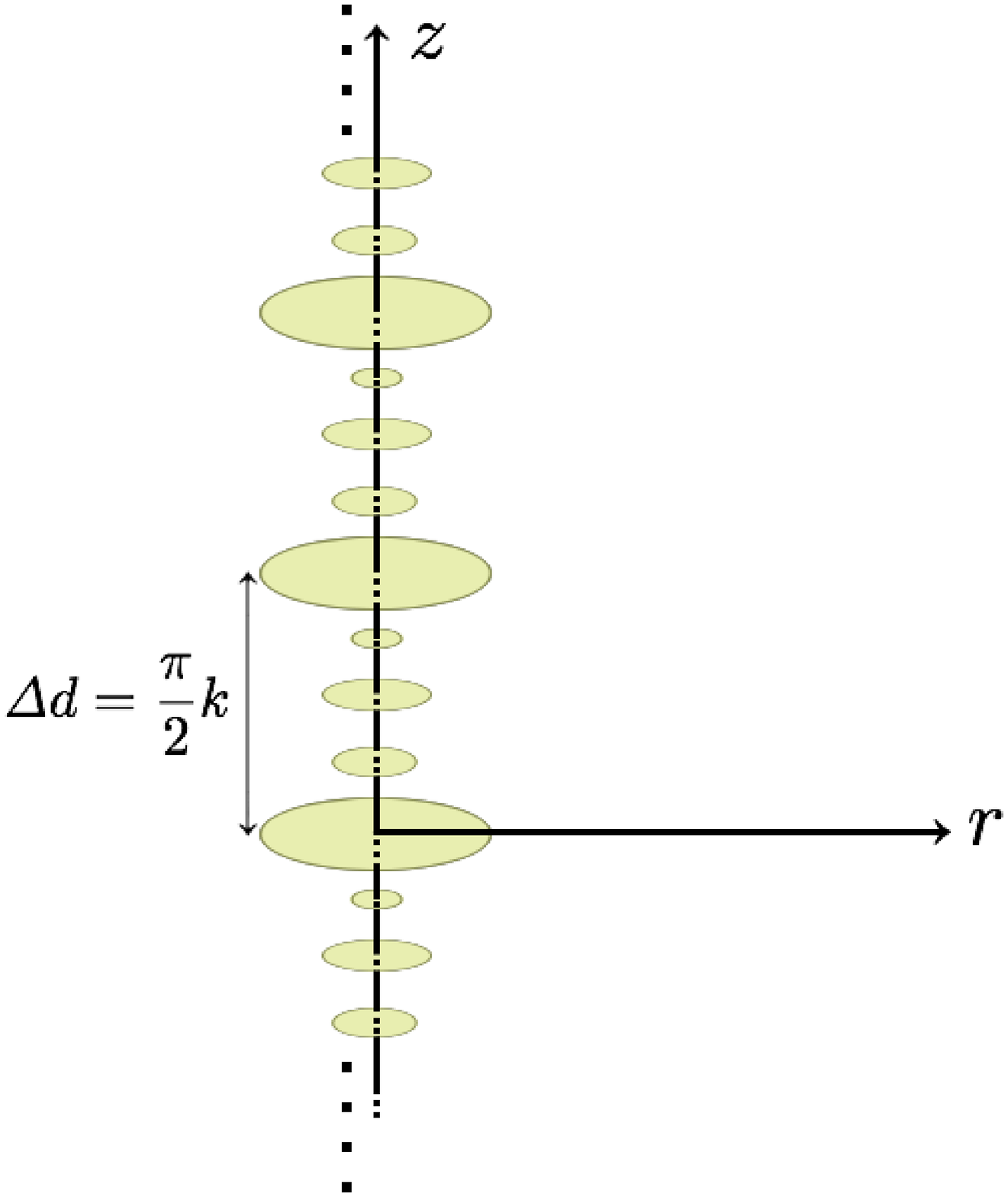}
\end{center}
\caption{Electrostatic systems corresponding to a general vacuum of SYM on $R\times S^2$ (left)
and SYM on $R\times S^3/Z_k$ (right). The conducting disks in the right figure are arranged periodically.}
\end{figure}

The condition that the charge densities vanish at the edges of the disks relates
the radii of the disks and the charges.
So the independent parameters of this solution are the total charges 
and the $z$-coordinates of the disks, as well as $V_0$ in \eqref{bg pot} or $W_0$ in \eqref{bg pot S2}.

While $V_0$ and $W_0$ are related to the magnitude of the dilaton,
the other parameters turn out to be related to the brane charges.
This geometry has an $S^2$ and an $S^5$ at each point on the $r$-$z$ plane.
One can show that on the $z$-axis 
the radius of $S^5$ becomes zero, 
while on the finite conducting disks, 
the radius of $S^2$ becomes zero.
So one can consider a various non-contractible 
$S^3$'s or $S^6$'s which are given by 
fibering the $S^2$ or $S^5$ on the path (on the $r$-$z$ plane) 
ending on different disks or on different 
segments of the $z$-axis, respectively.
On each $S^3$ or $S^6$,
one can measure the NSNS-flux or the RR-flux.
This relates the parameters of the electrostatic system 
to the brane charges.
As an example, let us consider the electrostatic system for PWMM with 
$\Lambda$ finite plates (see Fig. 1).
We denote the total charge, the radius and 
the $z$-coordinate of $s$-th disk
by $Q_s$, $R_s$ and $d_s$, respectively,
where $s=1,\cdots, \Lambda$.
In this case, there are $\Lambda$ independent 
non-contractible $S^3$'s and the same number of $S^6$'s in the geometry.
$Q_s$ and $d_s$ are related to the D2-brane charges $N_2^{(s)}$ and the NS5-brane charges $N_5^{(s)}$ as
\begin{align}
Q_s = \frac{\pi^2 N_2^{(s)}}{8} , \;\;\; 
d_s-d_{s-1} = \frac{\pi N_5^{(s)}}{2},
\label{D2 NS5 charge}
\end{align}
for $s=1, \cdots, \Lambda$. 
Here $d_0 =0$ denotes the position of the infinite plate.

We will show in section \ref{Emergent bubbling geometry} 
how the parameters of the electrostatic system are mapped to those of the corresponding gauge theory.

\subsection{Electrostatic problem for PWMM}
\label{electrostatic problem for PWMM}

In the following, we derive a Fredholm integral equation which determines
$\tilde{V}(r,z)$ in (\ref{Vbg+tildeV}).
We consider the situation shown in Fig.1  and denote 
the solution of $\tilde{V}(r,z)$ in the region $z \in [d_{s},d_{s+1})$
by $V_s (r,z)$, where $s$ runs from $0$ to $\Lambda$ and  
$(d_0, d_{\Lambda+1})=(0, \infty)$ is assumed.

We first note that a general solution to the Laplace equation 
in an axially symmetric system is given by $J_0(r u)e^{\pm zu}$
where $J_{0}(z)$ is the Bessel function of the first kind 
of order zero and $u$ is a positive real number.
So we can write $V_s (r,z)$ as 
\begin{align}
V_{s}(r,z)= \int^{\infty}_0 du (C_s(u)e^{zu}+D_s(u)e^{-zu})J_0(ru).
\end{align}
Now we have the boundary condition that
$\tilde{V}(r,z) \rightarrow 0 $ as $z \rightarrow 0$ 
and $z \rightarrow \infty$. This means 
\begin{align}
C_0 (u) = - D_0 (u) , \;\;\; C_{\Lambda}(u)=0.
\label{bc for cd}
\end{align}
We also have some continuation conditions for $V_{s}$'s at $z = d_s $. 
First, at $z =d_s$, $V_{s}$ should be equal to $V_{s-1}$. This is 
satisfied if
\begin{align}
C_s(u)e^{d_s u}+D_s(u)e^{-d_s u} =C_{s-1}(u)e^{d_s u}+D_{s-1}(u)e^{-d_s u}.
\label{first cond}
\end{align}
Second, when $z =d_s$ and $r >R_s $, not only $V_s$ but also 
$\frac{\partial V_s}{\partial z}$ should be continuous. So we have
\begin{align}
\int_0^{\infty} du u \left(
C_s(u)e^{d_s u}-D_s(u)e^{-d_s u} 
-C_{s-1}(u)e^{d_s u}+D_{s-1}(u)e^{-d_s u}
\right)J_0(ru) =0
\label{second cond}
\end{align}
for $r >R_s $.
Third, when $z =d_s$ and $r \leq R_s $ (i.e. on the conducting disk), 
the value of $V(r, z)$ should be constant: 
$V(r,d_s)=\Delta_s$. In terms of $C_s(u)$ 
and $D_s(u)$, this is written as
\begin{align}
\int_0^{\infty} du
\left(
C_s(u)e^{d_s u}+D_s(u)e^{-d_s u}
\right)J_0 (ru) = \Delta_s -V_{\rm b.g.}(r,d_s).
\label{third cond}
\end{align}
In order to solve the conditions (\ref{first cond}),
(\ref{second cond}) and (\ref{third cond}), 
we define 
\begin{align}
A_s (u)= u(C_s(u)-C_{s-1}(u))e^{d_s u}
-u(D_s(u)-D_{s-1}(u))e^{-d_s u}
\label{def of A}
\end{align}
for $s=1,2,\cdots, \Lambda$.
From (\ref{bc for cd}) and (\ref{first cond}), 
$C_s(u)$ and $D_s(u)$ can be written in terms of $A_s(u)$ as 
\begin{align}
&C_s(u) = -\sum_{t=s+1}^{\Lambda}\frac{e^{-d_t u}}{2u}A_t (u),
\nonumber\\
&D_s(u) = \sum_{t=1}^{\Lambda}\frac{e^{-d_t u}}{2u}A_t (u)
-\sum_{t=1}^{s}\frac{e^{d_t u}}{2u}A_t (u).
\label{cd in terms of a}
\end{align}
By substituting (\ref{def of A}) and (\ref{cd in terms of a}) to 
(\ref{second cond}) and (\ref{third cond}), we obtain
\begin{align}
&\int_0^{\infty} u^{-1} \sum_{t=1}^{\Lambda} 
(\delta_{st}+k_{st}(u))A_t(u)J_0(ru)du = F_s(r),
\;\; (0\leq r \leq R_s ) \nonumber \\ 
&\int_0^{\infty} A_s(u)J_0(ru)du =0, \;\; (R_s \leq r )
\label{equations for a}
\end{align}
where $k_{st}(u)$ and $F_s(r)$ are given by
\begin{align}
&k_{st}(u)=-e^{-(d_s+d_t)u}+(1-\delta_{st})e^{-|d_s-d_t|u},
\nonumber\\
&F_s(r)= -2(\Delta_s-V_{\rm b.g.}(r,d_s)).
\end{align}
As shown in appendix~\ref{dual integral equations},
the equations (\ref{equations for a})
can be reduced to the integral equations,
(\ref{dual eq for h2}) and (\ref{dual eq for h1}),
for the functions $h_s(u)$ defined by (\ref{def of hs})\footnote{
$n$, $I_{s1}$ and $I_{s2}$ in appendix~\ref{dual integral equations}
corresponds to 
$\Lambda$, $[0, R_s)$ and $[R_s,\infty)$ in our problem, 
respectively.}. 
For our problem, it is more convenient to work with the variables
\begin{align}
f_s(u) = -\frac{1}{4\sqrt{\pi}}uh_s(u).
\end{align}
Then, (\ref{dual eq for h1}) is written for $\{ f_s (x)\}$ as 
\begin{align}
f_{s}(x)+\frac{1}{\pi}\sum_{t=1}^{\Lambda} \int^{R_t}_{-R_t} du
\left[
-\frac{d_s+d_t}{(d_s+d_t)^2+(x-u)^2}+\frac{|d_s-d_t|}{(d_s-d_t)^2+(x-u)^2}
 \right]f_{t}(u) 
\nonumber\\ 
=\frac{1}{\pi}\left(
\Delta_s + \frac{2}{3}V_0 d_s^3 -2V_0 d_s x^2
\right),
\label{integral eq for fs}
\end{align}
and (\ref{dual eq for h2}) shows that $f_{s}$ is vanishing 
outside the region $[-R_s, R_s]$. 
Here we have defined $f_{s}(x)$ with negative $x$ as 
$f_{s}(x)=f_s(-x)$ and extended the domain to the entire real line.

The function $f_s (x)$ can be interpreted as the charge density 
on the $s$-th conducting disk as follows. 
For $z=d_s$ and $r\leq R_s$, we have
\begin{align}
\frac{\partial }{\partial z}V_s (r,d_s)-
\frac{\partial }{\partial z}V_{s-1} (r,d_s)
=K_{0,-\frac{1}{2}}h_s (r)
=4\int^{R_s}_r \frac{f_s'(u)}{\sqrt{u^2-r^2}} du.
\end{align}
On the other hand, this is equal to $-4\pi \sigma (r)$, where
$\sigma (r)$ is the charge density for the $r$ direction. 
Hence, the total charge on the disk can be computed as
\begin{align}
Q_s = -2 \int^{R_s}_0 dr \int^{R_s}_r du \frac{r f_s' (u)}{\sqrt{u^2 -r^2}}
= \int^{R_s}_{-R_s} du f_s(u). \label{charge}
\end{align}
These relations show that $f_s(u)$ corresponds to the charge 
density on the $s$-th plate projected onto a diameter direction.
These densities are fully determined by 
(\ref{integral eq for fs}) and so is the potential
which can be written in terms of $\{ f_s(u)\}$ as
\begin{align}
\tilde{V}(r,z)=
\sum_{s=1}^{\Lambda}
\int^{R_s}_{-R_s}dt
\left[
\frac{1}{\sqrt{(z-d_s+it)^2+r^2}}-
\frac{1}{\sqrt{(z+d_s+it)^2+r^2}}
\right]f_s(t).
\label{pot}
\end{align}
Note that $R_s$ and $\Delta_s$ are determined by $f_s(R_s)=0$ and \eqref{charge}.

\subsection{Electrostatic problem for $\mathcal{N}=8$ SYM on $R\times S^2$}
\label{electrostatic problem for SYM on RxS2}

The electrostatic system associated with the gravity dual of $\mathcal{N}=8$ SYM on $R\times S^2$ 
is shown in Fig.2 (left).
The case where $\Lambda=2$ and $R_1=R_2$ was studied in \cite{Ling:2006xi}. 
Here, we generalize their result.
It was shown in \cite{Ling:2006up,Ishiki:2006yr} that the solution for this system
can be obtained from the solution for PWMM by taking the D2-brane limit.
After the redefinitions 
$d_s\rightarrow d+d_s \; \left( 1\leq s\leq \Lambda \right), \; z\rightarrow d+z$,
D2-brane limit is written as
\begin{align}
 &d\to \infty ,\quad
 Q_s\; \text{: fixed},\quad
 V_0d=W_0\; \text{: fixed}.
\label{D2 limit}
\end{align}
Indeed, in this limit, Fig.1 becomes Fig.2 (left) and
the background potential for PWMM \eqref{bg pot} becomes
\begin{align}
 V_{\rm b.g.}(r,z)
 \to -W_0\left( \frac{2d^2}{3}+2dz\right)+W_0(r^2-2z^2).
 \label{D2 bg pot}
\end{align}
One can neglect the first term
since it does not affect the gravity solution 
which depends only on $\dot V$, $\ddot V$, $\dot V'$ and $V''$.
Thus, the background potential for PWMM \eqref{bg pot} exactly reduces to 
that for SYM on $R\times S^2$ \eqref{bg pot S2} in the limit \eqref{D2 limit}.

By taking  the D2-brane limit \eqref{D2 limit} of the integral equation \eqref{integral eq for fs}
and the potential \eqref{pot},
we obtain  
\begin{align}
& f_{s}(x)+\frac{1}{\pi}\sum_{t=1}^{\Lambda} \int^{R_t}_{-R_t} du
 \frac{|d_s-d_t|}{(d_s-d_t)^2+(x-u)^2}
 f_{t}(u) 
 =\frac{1}{\pi}\left(
 \Delta_s' + 2W_0d_s^2 -2W_0 x^2
 \right), 
\label{D2 integral eq}
\end{align}
and
\begin{align}
& \tilde{V}(r,z)=
 \sum_{s=1}^{\Lambda}
 \int^{R_s}_{-R_s}dt
 \frac{f_s(t)}{\sqrt{(z-d_s+it)^2+r^2}},
 \label{D2 pot}
\end{align}
respectively.
Here $\Delta_s'$ is a constant potential on the $s$-th disk, $V(r,d_s)=\Delta_s' \ (r<R_s)$.
Note that $R_s$ and $\Delta_s'$ are determined by $f_s(R_s)=0$ and \eqref{charge}.
The solution of \eqref{D2 integral eq} gives
a general solution to the electrostatic problem for SYM on $R\times S^2$.

\subsection{Electrostatic problem for $\mathcal{N}=4$ SYM on $R\times S^3/Z_k$}
\label{electrostatic problem for SYM on RxS3/Zk}

The electrostatic system associated with the gravity dual of $\mathcal{N}=4$ SYM on $R\times S^3/Z_k$ 
is shown in Fig.2 (right).
The case for the trivial vacuum was studied in \cite{Ling:2006xi}. 
Here, we generalize their result.
This can be obtained from that for SYM on $R\times S^2$ by compactifying the $z$ direction to $S^1$
with the background potential intact \cite{Ishiki:2006yr}.

We start from the solution for SYM on $R\times S^2$, \eqref{D2 integral eq} and \eqref{D2 pot},
with disks periodically arranged.
We change the labelling of the disks so that they are labelled by two integers $(s,\alpha)$, 
where $-\infty\leq s \leq \infty$ and $\alpha\in K\subset\{1,2,\cdots,k\}$.
$s$ is a label of a single period and $\alpha$ is that of each disk in the period.
So each period consists of $|K|$ conducting disks.
We put the position of each disk to be
\begin{align}
 d_{s,\alpha}=\frac{\pi}{2} \left(ks+\alpha -1\right).\quad
\end{align}
The charge $Q_{s,\alpha}$ and the radius $R_{s,\alpha}$ of each disk 
is independent of $s$: $Q_{s,\alpha}=Q_\alpha$ and $R_{s,\alpha}=R_\alpha$.
The charge density $f_{s,\alpha}(r)$ on each disk should also be independent of $s$:
\begin{align}
f_{s,\alpha}(r)=f_\alpha(r).
\end{align}

Note that the naive substitutions of these conditions to \eqref{D2 integral eq} and \eqref{D2 pot} do not make sense
because of the divergences coming from the periodicity.
As remarked in \cite{Ling:2006xi}, this divergence can be avoided by solving the electrostatic problem 
for the electric field rather than the potential.
Hence, by differentiating \eqref{D2 integral eq} with respect to $x$ and imposing the periodicity condition,
one can obtain the integral equations for the charge densities $\{f_\alpha(r)\}$,
\begin{align}
 f_\alpha '(x)+\sum_{\beta\in K}\int^{R_\beta}_{-R_\beta} du\,
 K_k\left(\frac{\alpha-\beta}{k},x,u\right)f_\beta '(u) 
 =-\frac{4}{\pi}W_0 x, \label{S3/Zk integral eq}
\end{align}
where
\begin{align}
 K_k(\nu,x,u)
& =\frac{1}{4\pi}\int_{-\infty}^{\infty} dp
\frac{\cosh \left\{ \frac{\pi k}{2} p\left(|\nu|-\frac{1}{2}\right)\right\}}{\sinh \frac{\pi k}{4} |p|}
\left(e^{i p(x-u)}-e^{ip(x+u)}\right).
\end{align}
The electric field generated by the conducting disks is obtained from \eqref{D2 pot} as
\begin{align}
E_r&=\sum_{s=-\infty}^{\infty}\sum_{\alpha\in K}\int _{-R_\alpha}^{R_\alpha}dt
 \frac{rf_\alpha(t)}{((z-2d_{s,\alpha}+it)^2+r^2)^{\frac{3}{2}}}, \n
E_z&=\sum_{s=-\infty}^{\infty}\sum_{\alpha\in K}\int _{-R_\alpha}^{R_\alpha}dt
 \frac{(z-2d_{s,\alpha}+it)f_\alpha(t)}{((z-2d_{s,\alpha}+it)^2+r^2)^{\frac{3}{2}}}.
\end{align}

\section{Emergent bubbling geometry in theories with $SU(2|4)$ symmetry}
\label{Emergent bubbling geometry}

In this section, we investigate PWMM around a general vacuum
as well as $\mathcal{N}=8$ SYM on $R\times S^2$ and $\mathcal{N}=4$ SYM on $R\times S^3/Z_k$.
First, we briefly review the results of PWMM obtained in \cite{Asano:2012zt}, in which
the localization was applied to a 1/4-BPS sector and a matrix integral that describes this sector was obtained.
We then show that the eigenvalue densities of the matrix integral 
satisfy the same integral equations as the charge densities of the corresponding electrostatic system.
We also show that the same relation holds for the other gauge theories.
Some properties of these theories are reviewed in appendix \ref{SYM on S2 S3/Zk}.

\subsection{Localization in PWMM}

We first review PWMM. We follow the notation used in \cite{Asano:2012zt}.
The action of PWMM is given in the ten-dimensional notation as\footnote{
Here, the time direction is assumed to be the Euclidean signature.},
\begin{align}
S&=\frac{1}{g^2}\int d\tau  \Tr\Bigl(
\frac{1}{4}F_{MN}F^{MN}
+\frac{m^2}{8}X_mX^m
+\frac{i}{2}\Psi \Gamma^M D_M \Psi
\Bigr),
\label{action of PWMM}
\end{align}
where
\begin{align}
F_{1M}&=D_1X_M=\partial_1X_M-i[X_1,X_M] \quad (M\neq 1),\n
F_{ab}&=m \varepsilon_{abc}X_{c}-i[X_{a},X_{b}], \quad
F_{am}=D_{a}X_m=-i[X_{a},X_m], \quad
F_{mn}=-i[X_m,X_n], \n
D_1\Psi&=\partial_1\Psi-i[X_1,\Psi], \quad
D_{a}\Psi=\frac{m}{8}\varepsilon_{abc}\Gamma^{bc}\Psi-i[X_{a},\Psi], \quad
D_m\Psi=-i[X_m,\Psi].
\label{F in PWMM}
\end{align}
The ranges of the indices are $M,N=1,\cdots,10$, 
$a,b=2,3,4$ and $m,n=5,\cdots,10$.
$X_1$ is the one-dimensional gauge field, 
$X_{a}$ and $X_m$ are $SO(3)$ and $SO(6)$ scalar fields and 
$\Psi$ is a 16 component fermionic field. 
In the following we put the mass parameter $m=2$.

The vacuum of PWMM is given by the fuzzy sphere solution, which is a matrix representation of $SU(2)$ algebra
\begin{align}
X_a=-2L_a=\bigoplus_{s=1}^{\Lambda}\bm{1}_{N_2^{(s)}}\otimes L_a^{[D_s]}.
\label{fuzzy sphere}
\end{align}
Any representation of $SU(2)$ is an $SU(2|4)$ symmetric vacuum.
The right-hand side of \eqref{fuzzy sphere} stands for the irreducible decomposition,
where $L_a^{[D]}$ are generators of $SU(2)$ algebra in $D$-dimensional irreducible representation,
$N_2^{(s)}$ are  multiplicities of $L_a^{[D_s]}$ 
and $\Lambda$ is the number of different irreducible representations.

The parameters of PWMM around the vacuum \eqref{fuzzy sphere} are identified with 
those of the electrostatic system in the gravity dual as \cite{LM}\footnote{
Note that $N_5^{(s)}$ is used in different meanings in the present paper and in \cite{Asano:2014vba}.}
\begin{align}
Q_s=\frac{\pi^2N_2^{(s)}}{8}, \quad
d_s=\frac{\pi}{2}D_s=\frac{\pi}{2}\sum_{t=1}^{s}N_5^{(t)}. \label{D d}
\end{align}
In addition, in \cite{Ling:2006up}, the case of $\Lambda =1$ was considered 
and $V_0$ was inferred to be related to the gauge coupling as
\begin{align}
V_0=\frac{h m^3}{8 g^2},
\label{V0}
\end{align}
where $h$ is a constant. 
In \cite{Asano:2014vba}, the constant $h$ was determined as 
\begin{align}
h=\frac{2}{\pi^2}. \label{h}
\end{align}
We claim that the relation (\ref{V0}) also holds for the theory around 
the general vacuum. 
In general, $V_0$ should be a function of $g^2$ and $\hat N$, 
where $\hat N=\sum_sN_2^{(s)}D_s$ is the total matrix size  \cite{Ling:2006up}.
Since we are interested in the region where the supergravity approximation is valid, $\hat N$ must be large. 
Now, let us consider the case where $D_s\gg D_1$ for any $s>1$.
Then, we are left with the electrostatic system made 
of an infinite conducting plate and a single conducting disk.
This is exactly the same situation studied in \cite{Ling:2006up}, so that 
$V_0$ should be given by (\ref{V0}). Thus, if $N_2^{(1)}D_1$ is large, 
the relation (\ref{V0}) holds. Since large $D_s$ means large $\hat N$, 
we conclude that $V_0(g^2, \hat N)=\frac{h m^3}{8 g^2}$ when $\hat N \gg 1$.

Since PWMM has one noncompact direction, 
in order to define the theory around a fixed vacuum precisely we have to specify the boundary condition.
Here we choose the boundary condition such that all fields approach to the vacuum configuration
as $\tau\rightarrow \pm \infty$. 
We also consider the 't Hooft limit, in which the tunneling between vacua are suppressed.
Thus, the path integral with these conditions correctly defines PWMM around the fixed vacuum.

In the following, we consider PWMM around a general fuzzy sphere vacuum given in \eqref{fuzzy sphere}.
In this theory we focus on the complex scalar
\begin{align}
\phi (\tau) = -X_4(\tau)+\sinh \tau X_9(\tau)+ i\cosh \tau X_{10}(\tau).
\label{phi PWMM}
\end{align}
$\phi$ is invariant under four supersymmetries (1/4-BPS) after the Wick-rotation along $X_{10}$.
From a symmetry argument, $\phi$ was found to describe the $r,z$ directions in \eqref{LM solution} 
\cite{Asano:2014vba}.
So, this 1/4-BPS sector is expected to correspond to the electrostatic system on the gravity side.
Applying the localization to operators made of $\phi$, we obtain \cite{Asano:2012zt}
\begin{align}
\langle \prod_{a} {\rm Tr}f_a(\phi (\tau_a)) \rangle
=\langle \prod_{a} {\rm Tr}f_a(2L_4+iM) \rangle_{MM}.
\label{result}
\end{align}
Here $M$ is a $\tau$-independent Hermitian matrix with the following block structure,
\begin{align}
M=\bigoplus_{s=1}^{\Lambda} ( M_s \otimes {\bf 1}_{D_s}),
\end{align}
where $M_s$ $(s=1,\cdots, \Lambda)$ are $N_2^{(s)}\times N_2^{(s)}$
Hermitian matrices.
In the right-hand side $\langle \cdots \rangle_{MM}$ stands for an expectation value 
with respect to the following partition function,
\begin{align}
&Z_{{\cal R}}=\int \prod_{s=1}^{\Lambda}
\prod_{i=1}^{N_2^{(s)}}dm_{si} Z_{\rm 1-loop}({\cal R},\{m_{si}\})
e^{-\frac{2}{g^2}\sum_{s}\sum_{i}D_sm_{si}^2},
\label{matrix model}
\end{align}
where ${\cal R}$ denotes the representation of (\ref{fuzzy sphere}),
$m_{si}$'s are eigenvalues of $M_s$ and
\begin{align}
Z_{\rm 1-loop}=
\prod_{s,t=1}^{\Lambda}
\prod_{J}
\prod_{i=1}^{N_2^{(s)}}\prod_{j=1}^{N_2^{(t)}}
\hspace{-5.5mm} {\phantom{\prod}}^{\prime}
\left[
\frac{\{(2J+2)^2+(m_{si}-m_{tj})^2\} \{(2J)^2+(m_{si}-m_{tj})^2\}}
{\{(2J+1)^2+(m_{si}-m_{tj})^2\}^2}
\right]^{\frac{1}{2}}.
\label{1loopdet}
\end{align}
In (\ref{1loopdet}), the product of $J$ runs from 
$|D_s-D_t |/2$ to $(D_s+D_t)/2-1$.
$\prod'$ represents that the second factor in the numerator with 
$s=t$, $J=0$ and $i=j$ is not included in this product.

\subsection{Localization in SYM on $R\times S^2$ and SYM on $R\times S^3/Z_k$}
\label{Localization S2 S3/Zk}

In this subsection, 
we show the results of the localization for SYM on $R\times S^2$ and SYM on $R\times S^3/Z_k$ 
around a general vacuum, which can be  obtained by applying the D2-brane limit 
and the T-duality to \eqref{matrix model} with \eqref{1loopdet} \cite{Ishiki:2006yr}.

\subsubsection*{SYM on $R\times S^2$}

SYM on $R\times S^2$ around a general monopole vacuum \eqref{vacuum of SYM on RxS2}
 can be obtained from PWMM in the D2-brane limit \eqref{D2 limit}.
The D2-brane limit in the PWMM language reads
\begin{align}
 &D_s= D+2q_s \quad \left( 1\leq s\leq \Lambda \right), \quad
 D\to \infty ,\quad
 N_2^{(s)}\; \text{: fixed},\quad
 \frac{D}{g^2}=\frac{4\pi}{m^2 g_{S^2}^2}\; \text{: fixed}.
\label{PWMM D2 limit}
\end{align}
In this limit, \eqref{matrix model} with \eqref{1loopdet} becomes, up to an overall constant,
\begin{align}
Z_{\{(q_s,N_2^{(s)})\}}^{R\times S^2}=
&\int \prod_{s=1}^{\Lambda} \prod_{i=1}^{N_2^{(s)}}dm_{si}
\prod_{s=1}^{\Lambda}\Delta(m_s)^2
\prod_{s=1}^{\Lambda} \prod_{i,j=1}^{N_2^{(s)}}
\left[
\frac{1+\left( \frac{m_{si}-m_{sj}}{2}  \right)^2}
{\{1+(m_{si}-m_{sj})^2\}^2}
\right]^{\frac{1}{2}}
\nonumber\\
&\times \prod_{s,t=1}^{\Lambda}
\prod_{\substack{J=|q_s-q_t| \\ J\neq 0}}^{\infty}
\prod_{i=1}^{N_2^{(s)}}
\prod_{j=1}^{N_2^{(t)}}
\left[
\frac{\left\{1+\left(\frac{m_{si}-m_{tj}}{2J+2}\right)^2\right\} 
\left\{1+\left(\frac{m_{si}-m_{tj}}{2J}\right)^2\right\}}
{\left\{1+\left(\frac{m_{si}-m_{tj}}{2J+1}\right)^2\right\}^2}
\right]^{\frac{1}{2}}
e^{-\frac{m\pi }{g_{S^2}^2}\sum_{s,i} m_{si}^2},
\label{matrix model S2}
\end{align}
where $\Delta(m_s)=\prod_{i<j}(m_{si}-m_{sj})$ is the Vandermonde determinant.
The parameters of SYM on $R\times S^2$ around \eqref{vacuum of SYM on RxS2}
 correspond to those in the gravity dual as
\begin{align}
Q_s=\frac{\pi^2N_2^{(s)}}{8}, \quad
d_s=\pi q_s, \quad
W_0=\frac{\pi^2 h m}{4g_{S^2}^2}.
\label{parameter relation}
\end{align}

\subsubsection*{SYM on $R\times S^3/Z_k$}

SYM on $R\times S^3/Z_k$ around a general vacuum characterized by the holonomy \eqref{holonomy}
can be obtained by applying the T-duality to SYM on $R\times S^2$.
The T-duality can be realized in the same manner as in section \ref{electrostatic problem for SYM on RxS3/Zk}.
In \eqref{matrix model S2}, we first replace $s\rightarrow (s,\alpha)$ where $-\infty \leq s\leq \infty$
and $\alpha\in K\subset \{1,2,\cdots,k \}$.
Next, we set
\begin{align}
2q_{s\alpha}=ks+\alpha-1, 
\end{align}
and then drop the $s$-dependence of $m_{s\alpha i}$ and $N_2^{(s\alpha)}$: 
$m_{s\alpha i}\rightarrow m_{\alpha i}$ and $N_2^{(s\alpha)}\rightarrow N_2^{(\alpha)}$.
Finally, up to an overall constant, we end up with
\begin{align}
Z_{\{(\alpha, N_2^{(\alpha)})\}}^{R\times S^3/Z_k}
=& \int  \prod_{\alpha}\prod_{i=1}^{N_{2}^{(\alpha)}} dm_{\alpha i}
\prod_\alpha
\Delta(m_\alpha)^2 \prod_{\alpha}\prod_{i,j=1}^{N_2^{(\alpha)}}
\left[ 
\frac{1+\left(\frac{m_{\alpha i}-m_{\alpha j}}{2}\right)^2 }{\{1+(m_{\alpha i}-m_{\alpha j})^2\}^2}
\right]^{\frac{1}{2}} 
\nonumber\\
&\times \prod_{u=-\infty}^{\infty}\prod_{\alpha,\beta}
\prod_{\substack{J=|ku/2+(\alpha-\beta)/2|\\J\neq 0}}^{\infty}
\prod_{i=1}^{N_2^{(\alpha)}}\prod_{j=1}^{N_2^{(\beta)}}
\left[
\frac{\left\{ 1+\left(\frac{m_{\alpha i}-m_{\beta j}}{2J+2} \right)^2   \right\}
\left\{ 1+\left(\frac{m_{\alpha i}-m_{\beta j}}{2J}\right)^2  \right\}}
{\left\{1+\left(\frac{m_{\alpha i}-m_{\beta j}}{2J+1}\right)^2\right\}^2}
\right]^{\frac{1}{2}} \n
&\times e^{-\frac{4\pi^2}{k g_{S^3/Z_k}^2}\sum_{\alpha, i}m_{\alpha i}^2},
\label{matrix model S3/Zk}
\end{align} 
where $\alpha, \beta$ run over the elements of $K$, $\Delta(m)=\prod_{i<j}(m_i-m_j)$ is the 
Vandermonde determinant. We have also used 
\begin{align}
\frac{1}{g_{S^2}^2}=\frac{4\pi}{m k g_{S^3/Z_k}^2}.
\end{align}

\subsection{Correspondence to the gravity side}

Now, we evaluate the matrix integral \eqref{matrix model} with \eqref{1loopdet} in the regime where
in the gravity side the classical gravity approximation is valid.
This regime corresponds in the gauge theory side to the 't Hooft limit
\begin{align}
N_2^{(s)}\rightarrow \infty, \quad \lambda^{(s)}=g^2N_2^{(s)}=\text{fixed},
\label{tHooft limit}
\end{align}
and 
\begin{align}
D_s-D_{s-1} \gg 1, \quad \lambda^{(s)}\gg D_s,
\label{limit}
\end{align}
for arbitrary $s$. 
The 't Hooft limit \eqref{tHooft limit} suppresses the bulk string coupling 
while the conditions \eqref{limit} make the $\alpha'$ corrections negligible.
Note that $D_s-D_{s-1}=\frac{\pi}{2}N_5^{(s)}$ is the square of the $S^5$ radius
in the $s$-th NS5-brane throat \cite{LM}.
Also, as shown in appendix C, 
the second condition of \eqref{limit} means that
the radius of $S^5$ near the tip of a disk in the electrostatic system is large.
In these limits, \eqref{tHooft limit} and \eqref{limit}, 
one can evaluate \eqref{result} by applying the saddle point approximation, which becomes exact in these limits.

When $D_s-D_{s-1} \gg 1$, one can rewrite the measure factor in \eqref{1loopdet} as 
\begin{align}
&\prod_{J=0}^{D_s-1}
\frac{\{(2J+2)^2+(m_{si}-m_{sj})^2\} \{(2J)^2+(m_{si}-m_{sj})^2\}}
{\{(2J+1)^2+(m_{si}-m_{sj})^2\}^2} \n
&=
 \tanh ^2\frac{\pi (m_{si}-m_{sj})}{2}
 \exp \left\{ \frac{2D_s}{(2D_s)^2+(m_{si}-m_{sj})^2}
 -\cdots \right\}
\end{align}
for $s=t$, and 
\begin{align}
&\prod_{J=|D_s-D_t|/2}^{(D_s+D_t)/2-1}
\frac{\{(2J+2)^2+(m_{si}-m_{tj})^2\} \{(2J)^2+(m_{si}-m_{tj})^2\}}
{\{(2J+1)^2+(m_{si}-m_{tj})^2\}^2} \n
&=
 \exp\left\{ \frac{D_s+D_t}{(D_s+D_t)^2+(m_{si}-m_{tj})^2}
 -\frac{|D_s-D_t|}{(D_s-D_t)^2+(m_{si}-m_{tj})^2}
 +\cdots\right\}
\end{align}
for $s\neq t$, where ``$\cdots$'' stands for $1/(D_s \pm D_t)$ corrections.
We introduce the eigenvalue densities defined for each $s$ as
\begin{align}
\rho^{(s)}(x)=\sum_{i=1}^{N_2^{(s)}}\delta(x-m_{si}).
\end{align}
In the large $N_2^{(s)}$ limit \eqref{limit}, $\rho^{(s)}(x)$'s become continuous functions.
Then, we obtain the effective action for \eqref{matrix model} with \eqref{1loopdet}
\begin{align}
S_{eff}
  &=\sum_{s=1}^{\Lambda} \frac{2D_s}{g^2} \int dx \ x^2 \rho^{(s)}(x)
  -\sum_{s=1}^{\Lambda}\frac{1}{2}  \int dxdy \ \log \tanh^2 \frac{\pi (x-y)}{2}\rho^{(s)}(x)\rho^{(s)}(y)
  \nonumber \\
  &
  -\sum_{s,t=1}^{\Lambda}\frac{1}{2} \int dx dy \left[
  \frac{D_s+D_t}{(D_s+D_t)^2+(x-y)^2}
  -\frac{|D_s-D_t|}{(D_s-D_t)^2+(x-y)^2}\right] \rho^{(s)}(x)\rho^{(t)}(y) \n
  &-\sum_{s=1}^{\Lambda}\mu_s\left(\int dx \rho^{(s)}(x)-N_2^{(s)}\right),
\label{effective action}
\end{align}
where $\mu_s$'s are the Lagrange multipliers for the normalization of $\rho^{(s)}(x)$'s.

We assume that $\rho^{(s)}(x)$ has its support on $[-x_m^{(s)},x_m^{(s)}]$. 
As shown in appendix C, in the limit of \eqref{tHooft limit} and \eqref{limit} 
the extents of $\rho^{(s)}(x)$ become large; $x_m^{(s)}\gg 1$. 
Using the fact that $x_m\log \tanh^2 \frac{\pi x_m y}{2}$ can be approximated to $-\pi \delta(y)$ 
as $x_m\rightarrow \infty$, we obtain the following saddle point equations
\begin{align}
&\rho^{(s)}(x) +\frac{1}{\pi}
\sum_{t=1}^{\Lambda}
 \int_{-x_m^{(t)}}^{x_m^{(t)}} du\left[ 
-\frac{D_s+D_t}{(D_s+D_t)^2+(x-u)^2}+\frac{|D_s-D_t|}{(D_s-D_t)^2+(x-u)^2}
 \right]\rho^{(t)}(u) \n
&=\frac{\mu_s}{\pi}-\frac{2D_s}{\pi g^2}x^2,
\label{saddle eq}
\end{align}
where $x_m^{(s)}$ and $\mu_s$ are determined from
\begin{align}
\rho^{(s)}(x_m^{(s)})=0 \quad \text{and} \quad \int_{-x_m^{(s)}}^{x_m^{(s)}}dx \rho^{(s)}(x)=N_2^{(s)}.
\label{condition of rho}
\end{align}

Notice that the saddle point equations of the eigenvalue 
densities \eqref{saddle eq} 
take a very similar form as the integral equations for the 
charge densities \eqref{integral eq for fs}.
In fact, by using the relations (\ref{D d}) and (\ref{V0}), 
one can find that they are exactly the same equations.
Thus, we arrive at the relations 
\begin{align}
g^2\rho^{(s)}(x)&=\frac{1}{V_0}\left(\frac{2}{\pi}\right)^3f_s\left(\frac{\pi}{2}x\right), \label{rho f}
\end{align}
and
\begin{align}
\frac{\pi}{2}x_m^{(s)}&=R_s. \label{x R}
\end{align}
Namely, the eigenvalue density on the gauge theory side 
has exactly the same functional form and parameter dependence as 
the charge density on the gravity side, up to the trivial rescaling.
Hence, they can naturally be identified with each other and 
this identification relates the degrees of freedom on the gauge theory side 
to the background geometry on the gravity side.
By integrating both sides of \eqref{rho f} over 
$[-x_m^{(s)},x_m^{(s)}]$ 
and using \eqref{charge} and \eqref{condition of rho},
we find that those relations are consistent with 
\eqref{V0} and \eqref{h}.

If one finds exact solutions of \eqref{integral eq for fs} and 
\eqref{saddle eq}, one can check the relations, 
\eqref{rho f} and \eqref{x R}, more explicitly.
Although we could not find general exact solutions, 
still we can solve those equations in particular parameter regions.
If a conducting disk is isolated at a distance from the other disks,
the term with an integration in the integral equation of the disk 
becomes negligible. Then, the solution is simply given by a 
quadratic function.
This is effectively the same situation as the D2-brane limit with 
$\Lambda =1$ considered in \cite{Asano:2014vba}. 
In the same way, we can consider two 
isolated disks which effectively form the same system 
as the NS5-brane limit with $\Lambda =1$ \cite{Asano:2014vba}. 
In these cases, one can find the exact solutions and 
check the relations, \eqref{x R} and \eqref{rho f}, 
more directly.

The equivalence between the charge density and the eigenvalue density 
also holds for the other gauge theories with $SU(2|4)$ symmetry.
As shown in section \ref{Localization S2 S3/Zk},
in the D2-brane limit, the partition function 
\eqref{matrix model} reduces to 
the matrix integral for SYM on $R\times S^2$ given by
\eqref{matrix model S2}.
In addition, by taking the T-duality, 
we end up with the matrix integral for
SYM on $R\times S^3/Z_k$ given by \eqref{matrix model S3/Zk}.
If we apply the corresponding limits to the integral equation 
\eqref{integral eq for fs} of the charge density, 
we obtain the integral equations 
\eqref{D2 integral eq} and \eqref{S3/Zk integral eq} for 
SYM on $R\times S^2$ and
SYM on $R\times S^3/Z_k$, respectively.
In these cases, the integral equations 
\eqref{D2 integral eq} and \eqref{S3/Zk integral eq} for the 
charge densities can also be identified with
the saddle point equations for \eqref{matrix model S2} and \eqref{matrix model S3/Zk}, respectively, 
where the same relations as 
\eqref{rho f} and \eqref{x R} hold.



\section{Summary}

In this paper, 
we showed that the bubbling geometries in type IIA supergravity 
are realized in the gauge theories with $SU(2|4)$ symmetry.
We found that the charge densities of the electrostatic systems in the gravity dual
are equivalent to the eigenvalue densities of the matrix integrals 
which govern the 1/4-BPS sector of the gauge theories.

On the gravity side, the bubbling geometries are 
given in terms of the electrostatic potential of electrostatic systems with conducting disks.
First, we have considered the electrostatic system corresponding to PWMM around a general vacuum.
We have shown that the boundary conditions of the potential are given by a system of dual integral equations.
Extending the method to analyze the dual integral equations written in \cite{Sneddon}, 
we have reduced the dual integral equations to 
the Fredholm integral equations of the second kind for the charge densities on the disks.
By taking the D2-brane limit or performing the T-duality as well as the D2-brane limit, 
we have also obtained the same type of integral equations for the charge densities
in the electrostatic system corresponding to SYM on $R\times S^2$ or SYM on $R\times S^3/Z_k$.

On the gauge theory side, we have investigated the matrix integrals that describe 1/4-BPS sectors of
the gauge theories. First, we have considered the case for PWMM around a general vacuum in the regime 
where the supergravity approximation is valid.
In this regime, we have derived the saddle point equations of the eigenvalue densities of the matrix integral,
which are almost the same integral equations for the charge densities on the gravity side.
Then we have found that under the identifications of \eqref{D d} and \eqref{V0}
the integral equations of the eigenvalue densities are exactly equivalent to those for the charge densities.
As the D2-brane limit and the T-duality of PWMM lead to the other gauge theories with $SU(2|4)$ symmetry, 
that is, SYM on $R\times S^2$ and SYM on $R\times S^3/Z_k$,
we have also shown the equivalence of the charge densities and the eigenvalue densities.
Thus, we have concluded that since the bubbling geometries are completely determined by the charge densities
the geometries are constructed from the eigenvalue densities of the gauge theories with $SU(2|4)$ symmetry.

Finally, let us comment on IIA LST on $R\times S^5$,
which is another theory with $SU(2|4)$ symmetry. 
Like other $SU(2|4)$ symmetric theories, 
LST on $R\times S^5$ is thought to have many discrete vacua 
and for each vacuum there exists a gravity dual given by type IIA bubbling geometry \cite{LM}.
The gravity dual of LST around the trivial vacuum was elaborated in \cite{Ling:2006up} 
and shown to be obtained from a double scaling limit of the gravity dual of PWMM around a particular vacuum.
Although it is also expected that the gravity dual of LST around a general vacuum can be obtained 
from the same kind of double scaling limit of the gravity dual of PWMM, some careful analysis seems to be needed.
So, we will return to this issue in a separate paper.

\section*{Acknowledgments}
The work of Y.A. and S.S. is supported in part by the JSPS Research Fellowship for Young Scientists.
The work of S.S is also supported in part by the JSPS Grant-in-Aid for Scientific Research (A) No. 23244057.

\appendix
\section{Dual integral equations}
\label{dual integral equations}
Let $A(u)$ be a function defined on $[0,\infty)$.
We define two functions on $[0,\infty)$ using $A(u)$ as
\begin{align}
\phi(x)&=\int^{\infty}_0 u^{-2 \alpha}(1+k(u))A(u)J_{\nu}(xu)du, 
\label{phi} \\
\chi(x)&=\int^{\infty}_0A(u)J_{\nu}(xu)du, 
\label{chi}
\end{align}
where $\alpha$ is a positive half integer, $k(u)$ is a given weight 
function on $[0,\infty)$ and $J_{\nu}(z)$ is 
the Bessel function of the first kind of order ${\nu}$ defined by
\begin{align}
J_{\nu}(z)= \sum_{s=0}^{\infty}\frac{(-1)^{s}(z/2)^{\nu+2s}}
{s! \Gamma(\nu+s+1)}.
\end{align}
In this appendix, we consider a problem of finding a solution for 
$A(u)$, which solves the following equations called the dual integral equations.
\begin{align}
\phi(x) =F(x)  \;\;\; {\rm for} \;\; x \in I_1.
\label{phi F}
\\
\chi(x) =G(x)  \;\;\; {\rm for} \;\; x \in I_2.
\label{chi G}
\end{align}
Here we have divided the positive real line $[0,\infty)$ to two segments 
denoted by $I_1$ and $I_2$, which are written as $[0,c)$ and $[c,\infty)$, 
respectively.
$F(x)$ and $G(x)$ are assumed to be known functions 
defined on $I_1$ and $I_2$, respectively.
We will see that the problem reduces to a problem of solving 
a single Fredholm integral equation of the second kind \cite{Sneddon}.

In the following, for any function $f(x)$ on $[0,\infty)$, 
we denote by $f_1(x)$ and $f_2(x)$ the restrictions of $f(x)$ to 
$I_1$ and $I_2$, respectively. 
We assume that $f_i(x)=0$ unless $x \in I_i $ ($i=1,2$),
so that the original function can be written as $f(x)=f_1(x)+f_2(x)$. 
For example, the equation (\ref{phi F}) can be written in this notation 
as $\phi_1(x) = F(x)$ for $x \in I_1$.

We also introduce the modified Hankel transformation,
\begin{align}
S_{\eta, \alpha} f(x) := \left( \frac{2}{x} \right)^{\alpha}
\int^{\infty}_0 t^{1-\alpha} f(t)J_{2\eta+\alpha}(xt)dt.
\end{align}
The inverse transformation is given by 
\begin{align}
S_{\eta,\alpha}^{-1}=S_{\eta+\alpha,-\alpha}.
\label{inverse Hankel}
\end{align}
It is easy to see that
\begin{align}
&\phi (x) = 
\left( \frac{x}{2} \right)^{\alpha}
S_{\nu /2-\alpha, 2\alpha}\{(1+k)\cdot \psi \} (x), 
\label{phi 2}\\
&\chi(x) = S_{\nu /2, 0} \psi (x) , 
\label{chi 2}
\end{align}
where $\psi (u)$ is defined by
\begin{align}
A(u) = u\psi (u),
\end{align}
and the dot in (\ref{phi 2}) denotes 
the product of functions defined as usual by
$f\cdot g (x) =f(x)g(x)$.

We first put 
\begin{align}
\psi (u)= S_{\nu/2, -\alpha}h(u),
\label{psi to h}
\end{align}
and substitute this to (\ref{chi 2}).
Then, after performing the inverse 
transformation (\ref{inverse Hankel})
twice, we obtain
\begin{align}
h(x)=S_{\nu/2 +\alpha, -\alpha}
S_{\nu/2, 0}\chi (x) =K_{\nu /2+\alpha, -\alpha}\chi (x),
\label{h and chi}
\end{align}
where we have defined
\begin{align}
K_{\eta,\alpha+\beta}=S_{\eta, \alpha}S_{\eta+\alpha, \beta}.
\end{align}
One can show that the transformation $K_{\eta,\alpha}$ can be written as
\begin{align}
K_{\eta,\alpha}f(x)
=
\left\{
\begin{array}{l}
\frac{2 x^{2\eta}}{\Gamma(\alpha)}
\int^{\infty}_x
(u^2-x^2)^{\alpha-1}u^{-2\alpha -2 \eta +1}
f(u)du   \;\;\; {\rm for} \;\; 0< \alpha,  \\
-\frac{x^{2\eta -1}}{\Gamma(1+\alpha )}
\frac{d}{dx} \int^{\infty}_{x}u^{-2\alpha -2\eta +1} (u^2-x^2)^{\alpha}
f(u)du   \;\;\; {\rm for} \;\; -1< \alpha <0.
\end{array}
\right.
\label{def of K}
\end{align}
See \cite{Sneddon} for the definition for $\alpha < -1$.
From (\ref{h and chi}) and (\ref{def of K}), we find that 
the solution for $h_2(x)$ is given by
\begin{align}
h_2(x)=K_{\nu/2 +\alpha, -\alpha }\chi_2(x) = 
K_{\nu/2 +\alpha, -\alpha }G(x).
\label{h2 and G}
\end{align}
Similarly, by applying the same calculation to (\ref{phi 2}),
one can obtain the following equation for $x \in I_1 $.
\begin{align}
h_1(x)+S_{\nu /2-\alpha,\alpha}
\{k \cdot S_{\nu /2, -\alpha} h_1 \}(x) =H(x).
\label{eq for h1}
\end{align}
The function $H(x)$ is defined by\footnote{The last term in (\ref{def of H})
is obtained by using (\ref{h2 and G}) and
the relation,
$S_{\eta,\alpha}K_{\eta+\alpha, \beta} = S_{\eta, \alpha+\beta}$.}
\begin{align}
H(x) = \left( \frac{2}{x} \right)^{2\alpha}
I_{\nu /2, -\alpha} F(x)
-S_{\nu /2 -\alpha, \alpha }k(x) S_{\nu /2,0}G(x),
\label{def of H}
\end{align}
where $I_{\eta, \alpha+\beta}$ is defined by
\begin{align}
I_{\eta, \alpha+\beta}=S_{\eta+\alpha, \beta} S_{\eta, \alpha},
\end{align}
and it is written more explicitly as
\begin{align}
I_{\eta,\alpha}f(x)
=
\left\{
\begin{array}{l}
\frac{2x^{-2\alpha -2\eta}}{\Gamma(\alpha)}
\int^x_0 u^{2\eta +1} (x^2-u^2)^{\alpha-1}f(u) du
\;\;\; {\rm for} \;\; 0< \alpha,  \\
\frac{x^{-2\alpha -2\eta-1}}{\Gamma(1+\alpha)}
\frac{d}{dx}
\int^x_0 u^{2\eta +1} (x^2-u^2)^{\alpha }f(u) du
\;\;\; {\rm for} \;\; -1< \alpha <0.  \\
\end{array}
\right.
\end{align}
Note that $H(x)$ depends only on the known functions 
$k(x)$, $F(x)$ and $G(x)$.
For the second term in the left-hand side of (\ref{eq for h1}),
we interchange the order of the integration, so that
\begin{align}
S_{\nu /2-\alpha,\alpha}
\{k \cdot S_{\nu /2, -\alpha} h_1 \}(x)
= \int^{1}_0 K(x,u)h_1(u) du,
\end{align}
where the integral kernel $K(x,u)$ is defined by
\begin{align}
K(x,u)= u\left(\frac{u}{x} \right)^{\alpha}
\int^{\infty}_{0} t k(t) J_{\nu-\alpha}(xt)J_{\nu -\alpha} (ut) dt.
\end{align}
Hence, we conclude that $h_1$ is the solution of the Fredholm integral 
equation of the second kind,
\begin{align}
h_1(x) +\int^{1}_0 K(x,u)h_1(u) du = H(x).
\label{Fredholm}
\end{align}
The two equations (\ref{h2 and G}) and (\ref{Fredholm})
fully determines $h(x)=h_1(x)+h_2(x)$.

Now, we consider a generalization of this result to 
vector-valued functions. 
This is easy since the above problem is linear in $A(u)$. 
Let us consider a set of functions $\{ A_s(u)| s=1,\cdots,n  \}$
which are determined by the equations,
\begin{align}
\phi_s(x) =F_s(x)  \;\;\; {\rm for} \;\; x \in I_1^{(s)},
\label{phi F general}
\\
\chi_s(x) =G_s(x)  \;\;\; {\rm for} \;\; x \in I_2^{(s)},
\label{chi G general}
\end{align}
where $\phi_s(x) $ and $\chi_s(x)$ are now defined by
\begin{align}
\phi_s(x)&=\int^{\infty}_0 u^{-2 \alpha}
\sum_{t=1}^{n}(\delta_{st}+k_{st}(u))A_t(u)J_{\nu}(xu)du, 
\label{phi general} \\
\chi_s(x)&=\int^{\infty}_0A_s (u)J_{\nu}(xu)du.
\label{chi general}
\end{align}
For each $s$, $F_s(x)$ and $G_s(x)$ are assumed to be 
known functions and $I_{1,2}^{(s)}$ are the two connected 
intervals, the sum of which is equal to $[0,\infty)$.
If we write for each $s$
\begin{align}
A_s(u)= u S_{\nu/2, -\alpha}h_s(u),
\label{def of hs}
\end{align}
it is easy to see that 
$h_s(u)$ are determined by the following equations.
\begin{align}
&h_{s2}(x)= K_{\nu/2 +\alpha, -\alpha }G_s(x).
\label{dual eq for h2}
\\
&h_{s1}(x) +\sum_{t=1}^n \int^{1}_0K_{st}(x,u) h_{t1}(u) du = H_s(x),
\label{dual eq for h1}
\end{align}
where $H_s(x)$ and $K_{st}(x,y)$ are defined by
\begin{align}
H_s(x) = \left( \frac{2}{x} \right)^{2\alpha}
I_{\nu /2, -\alpha} F_s(x)
-\sum_{t=1}^{n} S_{\nu /2 -\alpha, \alpha }k_{st}(x) S_{\nu /2,0}G_t(x),
\label{def of H general}
\end{align}
and 
\begin{align}
K_{st}(x,u)= u\left(\frac{u}{x} \right)^{\alpha}
\int^{\infty}_{0} t k_{st}(t) J_{\nu-\alpha}(xt)J_{\nu -\alpha} (ut) dt.
\end{align}

\section{SYM on $R\times S^2$ and SYM on $R\times S^3/Z_k$}
\label{SYM on S2 S3/Zk}

In this appendix, we review SYM on $R\times S^2$ and SYM on $R\times S^3/Z_k$.
See for detail \cite{Asano:2012zt,Ishiki:2006yr}.

\subsubsection*{SYM on $R\times S^2$}

The action of SYM on $R\times S^2$ is given by
\begin{align}
S_{R\times S^2}&=\frac{1}{g_{S^2}^2}\int d\tau d\Omega_2 \Tr\Bigl(
-\frac{1}{2}(\partial_1 X_{b}-iL^{(0)}_{b}X_1-i[X_1,X_{b}])^2-\frac{1}{2}(D_1X_m)^2 \n
&\hspace{35mm}
-\frac{1}{4}(2\varepsilon_{abc}X_{c}+iL^{(0)}_{a}X_{b}-iL^{(0)}_{b}X_{a}
-i[X_{a},X_{b}])^2 \n
&\hspace{35mm}
+\frac{1}{2}(L_a^{(0)}X_m-[X_a,X_m])^2-\frac{1}{2}X_m^2+\frac{1}{4}[X_m,X_n]^2 \n
&\hspace{35mm}
-\frac{i}{2}\Psi \Gamma^1 \partial_1 \Psi
+\frac{1}{2}\Psi\Gamma^{a} L^{(0)}_{a} \Psi-\frac{3i}{8}\Psi\Gamma^{234}\Psi 
-\frac{1}{2}\Psi\Gamma^M[X_M,\Psi]
\Bigr), \label{SYM on RxS2 1}
\end{align}
where $L_{a}^{(0)}$ are ordinary angular momentum operators and
the radius of $S^2$ is set to be $\frac{1}{2}$ \cite{Asano:2012zt,Ishiki:2006yr}.
$X_1$ is the gauge field corresponding to the $R$ direction, 
$X_m$ are the scalar fields and $\Psi$ is a ten-dimensional Majorana-Weyl spinor with 16 components.
$X_{a}$ include gauge fields and a scalar field:
\begin{align}
\vec{X}=\Phi \vec{e}_r +a_{2} \vec{e}_\varphi-a_{3}\vec{e}_\theta,
\label{def of Phi}
\end{align}
where $\vec{X}=(X_2,X_3,X_4)$, 
$\vec{e}_r=(\sin\theta\cos\varphi,\sin\theta\sin\varphi,\cos\theta)$, 
$\vec{e}_\theta=(\cos\theta\cos\varphi,\cos\theta\sin\varphi,-\sin\theta)$ and
$\vec{e}_\varphi=(-\sin\varphi,\cos\varphi,0)$.
$a_{2}$ and $a_{3}$ are the gauge fields in the local Lorentz frame and $\Phi$ is the scalar field.
The vacuum of this theory is given by the Dirac monopole
\begin{align}
&\hat{a}_2=0,\quad \hat{a}_3=-\frac{\cos\theta\mp 1}{\sin\theta}\hat{\Phi},\n
&\hat{\Phi}=2 \ \diag(q_{1}\bm{1}_{N_{1}},\cdots, 
q_{s} \bm{1}_{N_{s}},\cdots,
q_{\Lambda}\bm{1}_{N_{\Lambda}}),
\label{vacuum of SYM on RxS2}
\end{align}
where $q_s\in Z/2$ are monopole charges. The upper and lower signs represent 
the northern region where $0\leq \theta < \pi$
and the southern region where $0<\theta\leq \pi$ on $S^2$, respectively.

The localization can be applied to the following scalar field sitting at $(\tau,\theta,\varphi)=(\tau,0,0)$,
\begin{align}
\phi (\tau,0,0) = -X_4(\tau,0,0)+\sinh \tau X_9(\tau,0,0)+ i\cosh \tau X_{10}(\tau,0,0).
\label{phi S2}
\end{align}
This scalar field preserves the same supersymmetries as \eqref{phi PWMM}.
Note that $X_4(\tau,0,0)=\Phi(\tau,0,0)$.
The vev of operators made of \eqref{phi S2} can be computed in terms of the matrix integral
obtained in the D2-brane limit of \eqref{matrix model} with \eqref{1loopdet}.

\subsubsection*{SYM on $R\times S^3/Z_k$}

$S^3/Z_k$ is the $Z_k$-orbifold of the round $S^3$ along the $S^1$-fiber direction.
When $S^3$ is parametrized by the coordinates $(\theta,\varphi,\psi)$
($0\leq \theta \leq \pi, \ 0\leq \varphi\leq 2\pi, \ 0\leq \psi \leq 4\pi$)
with the metric
\begin{align}
ds^2_{S^3}=\frac{1}{4}\left\{d\theta^2+\sin^2\theta d\varphi^2+(d\psi+\cos\theta d\varphi)^2\right\},
\end{align}
$S^3/Z_k$ is realized as the following coordinate identification; 
$(\theta, \varphi,\psi)\sim (\theta,\varphi,\psi+4\pi/k)$.
Then, SYM on $R\times S^3/Z_k$ is obtained by keeping the modes of fields of SYM on $R\times S^3$ 
that are invariant under the $Z_k$ action.

The action of SYM on $R\times S^3/Z_k$ takes the form,
\begin{align}
S_{R\times S^3/Z_k} 
&=\frac{1}{g_{S^3/Z_k}^2}\int d\tau d\Omega_{S^3/Z_k} \Tr\Bigl[
-\frac{1}{2}(\partial_1 X_{b}-i\cL_{b}X_1-i[X_1,X_{b}])^2-\frac{1}{2}(D_1X_m)^2 \n
&\hspace{35mm}
-\frac{1}{4}(2\varepsilon_{abc}X_{c}+i\cL_{a}X_{b}-i\cL_{b}X_{a}-i[X_{a},X_{b}])^2 \n
&\hspace{35mm}
+\frac{1}{2}(\cL_aX_m-[X_a,X_m])^2-\frac{1}{2}X_m^2+\frac{1}{4}[X_m,X_n][X^m,X^n] \n
&\hspace{35mm}
-\frac{i}{2}\Psi \Gamma^1 \partial_1 \Psi
+\frac{1}{2}\Psi\Gamma^{a} \cL_{a} \Psi-\frac{3i}{8}\Psi\Gamma^{234}\Psi-\frac{1}{2}\Psi\Gamma^M[X_M,\Psi]
\Bigr],
\label{SYM on RxS3/Zk}
\end{align}
where $\cL_{a}$ are the Killing vectors on $S^3/Z_k$ \cite{Asano:2012zt,Ishiki:2006yr}.
$X_1$ and $X_{a}$ are the gauge fields for $R$ and $S^3/Z_k$ directions, respectively, 
$X_m$ are the scalar fields and $\Psi$ is a ten-dimensional Majorana-Weyl spinor with 16 components.

The vacuum of this theory is parametrized by the holonomy around the $S^1$-fiber direction of $S^3/Z_k$.
The holonomy can be labeled by
\begin{align}
\{(e^{2\pi i\alpha/k},N_2^{(\alpha)})\}_{\alpha \in K},
\label{holonomy}
\end{align}
where $K\subset \{1,2,\cdots,k\}$ 
and $N_2^{(\alpha)}$ are the multiplicities of the holonomy phase $e^{2\pi i \alpha/k}$.
The total matrix size is given by $N_2=\sum_{\alpha\in K}N_2^{(\alpha)}$.
Note that when $k=1$ this theory is the SYM on $R\times S^3$ and has the unique and trivial vacuum.

The localization can be applied to the following Wilson loop operator defined at a fixed $\tau$,
\begin{align}
W(\tau)=\frac{1}{N_2}\mathrm{Tr}P \exp\left(-2\pi i \int_0^{1}ds 
\left\{-X_4(x(s))+\sinh \tau X_9(x(s))+i\cosh \tau X_{10}(x(s))\right\}\right),
\label{W}
\end{align}
where $x^\mu(s)=(\tau,0,0,4\pi s)$.
This also preserves the same supersymmetries as \eqref{phi PWMM}. 
The correlation functions of \eqref{W} can be computed in terms of the matrix integral 
obtained by applying the T-duality to \eqref{matrix model S2}.

\section{Condition for large $S^5$ radius}
In this appendix, we show that $\lambda^{(s)}\gg D_s$ is a sufficient 
condition for the large $S^5$ radius
at the tips of the disks in the electrostatic problem.
At the tip of a disk, the disk radius $R$ and the radius 
$R_{S^5}$ of $S^5$ are related as \cite{Ling:2006up}
\begin{align}
R = \frac{R^{2}_{S^5}}{4}
\end{align}
in the string unit, $\alpha'=1$.
Then, under the identification (\ref{x R}), 
the $S^5$ radius is large if and only if $x_m^{(s)}\gg 1$.
In the following, we show that $x_m^{(s)}\gg 1$ if 
$\lambda^{(s)}\gg D_s$.
We assume that the index $s$ labels the disks in the 
order of the $z$-coordinate, namely, 
$D_{s-1}<D_s$ ($s=1,2,\cdots,\Lambda$).

First we divide the theory described by
(\ref{effective action}) into three parts.
The first is the free part, the action of which is given by
\begin{align}
S_{1}= \sum_{s=1}^{\Lambda}
\int dx
\left( \frac{2D_s}{g^2}
(x^2-\mu_s) \rho^{(s)}(x) 
+ \frac{\pi}{2} (\rho^{(s)}(x))^2
\right).
\end{align}
The second is the self-interaction part given by 
\begin{align}
S_{2}=- \frac{1}{2}
\sum_{s=1}^{\Lambda} \int dx dy
\frac{2D_s}{(2D_s)^2 +(x-y)^2} \rho^{(s)}(x)\rho^{(s)}(y).
\label{s2}
\end{align}
The third is the interaction between different $s$ and $t$, 
defined by
\begin{align}
S_{3}=- \frac{1}{2}
\sum_{s\neq t} \int dx dy
\left[
\frac{D_s+D_t}{(D_s+D_t)^2 +(x-y)^2}-
\frac{|D_s-D_t|}{(D_s-D_t)^2 +(x-y)^2}
\right]
 \rho^{(s)}(x)\rho^{(t)}(y).
\label{s3}
\end{align}
The total theory is described by the sum of these. 
But for the moment, let us consider more generally the theory 
defined by $S(\alpha,\beta)=S_1 +\alpha S_2 +\beta S_{3} $, 
where $\alpha$ and $\beta$ are parameters.
We start with the simplest free theory with $\alpha= \beta =0$. 
In this case, the extents of the eigenvalues can be easily estimated as
$x_m^{(s)}\sim (\lambda^{(s)}/D_s)^{1/3}$.
This gives a typical length scale of the free theory.
From (\ref{s2}) and (\ref{s3}), one can also read off the typical length 
scale of the interaction potentials. 
For the self-interaction, it is given by $\Delta x \sim D_s$,
where $\Delta x $ denotes the separation distance 
between two eigenvalues. 
For the interaction between different $s$ and $t$, the 
scale (for a fixed $s$) is equal or greater than $D_s^{3/4}$, 
namely, $\Delta x \gtrsim D_s^{3/4}$.
The lower bound is saturated by the interaction between 
$s$ and $t=s\pm 1$.
Then, let us consider turning on the interactions
to recover the theory with $\alpha = \beta =1$. 
The typical scale $(\lambda^{(s)}/D_s)^{1/3}$ of the free theory should 
be modified by the interactions, which have structures with the 
length scale equal or greater than $D_s^{3/4}$ (Note that we always 
assume that $D_s \gg 1$). 
The modified scale should be at least greater than 
${\rm min}((\lambda^{(s)}/D_s)^{1/3}, D_s^{3/4})$, since there is nothing 
which provides a finer scale than these.
If the modified scale is $x_m^{(s)} \sim D_s^{3/4}$, this is always 
large enough when $D_s \gg 1$. If the modified scale is 
$x_m^{(s)} \sim (\lambda^{(s)}/D_s)^{1/3}$, this is large if 
$\lambda^{(s)} \gg D_s $. Therefore, we conclude that 
if $\lambda^{(s)} \gg D_s $, 
the typical extents of the eigenvalues are always much greater than 1.

\end{document}